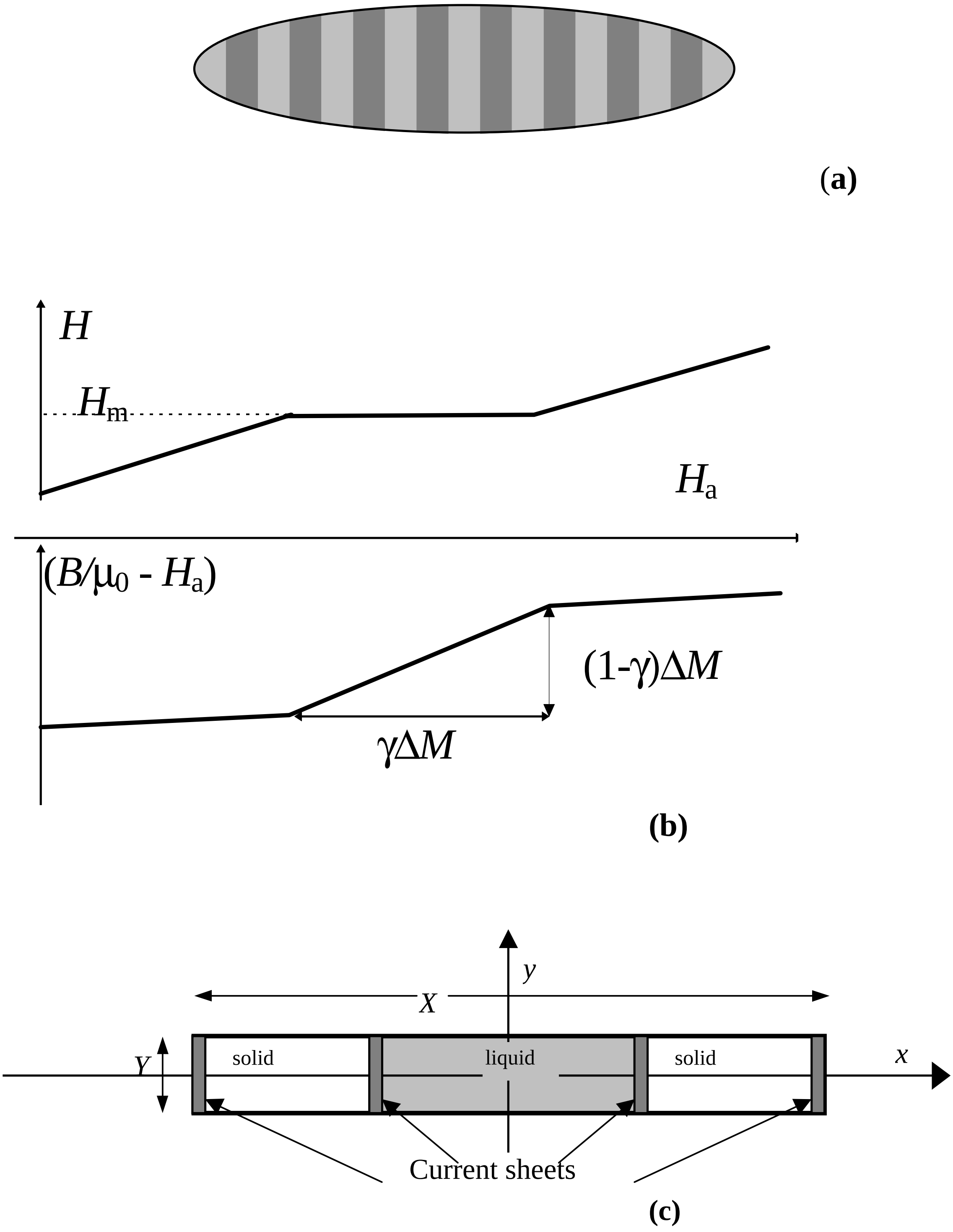

Figure 1 An intermediate state consisting of alternate regions of solid and liquid is illustrated in (a) for the case of an ellipsoidal specimen with a field applied along the short axis. The predicted variation of $H$ and $B$ with applied field $H_a$ as the ellipsoidal specimen passes through the intermediate state are shown in (b). The current-sheet model used to study the behaviour of rectangular specimens is illustrated in (c).

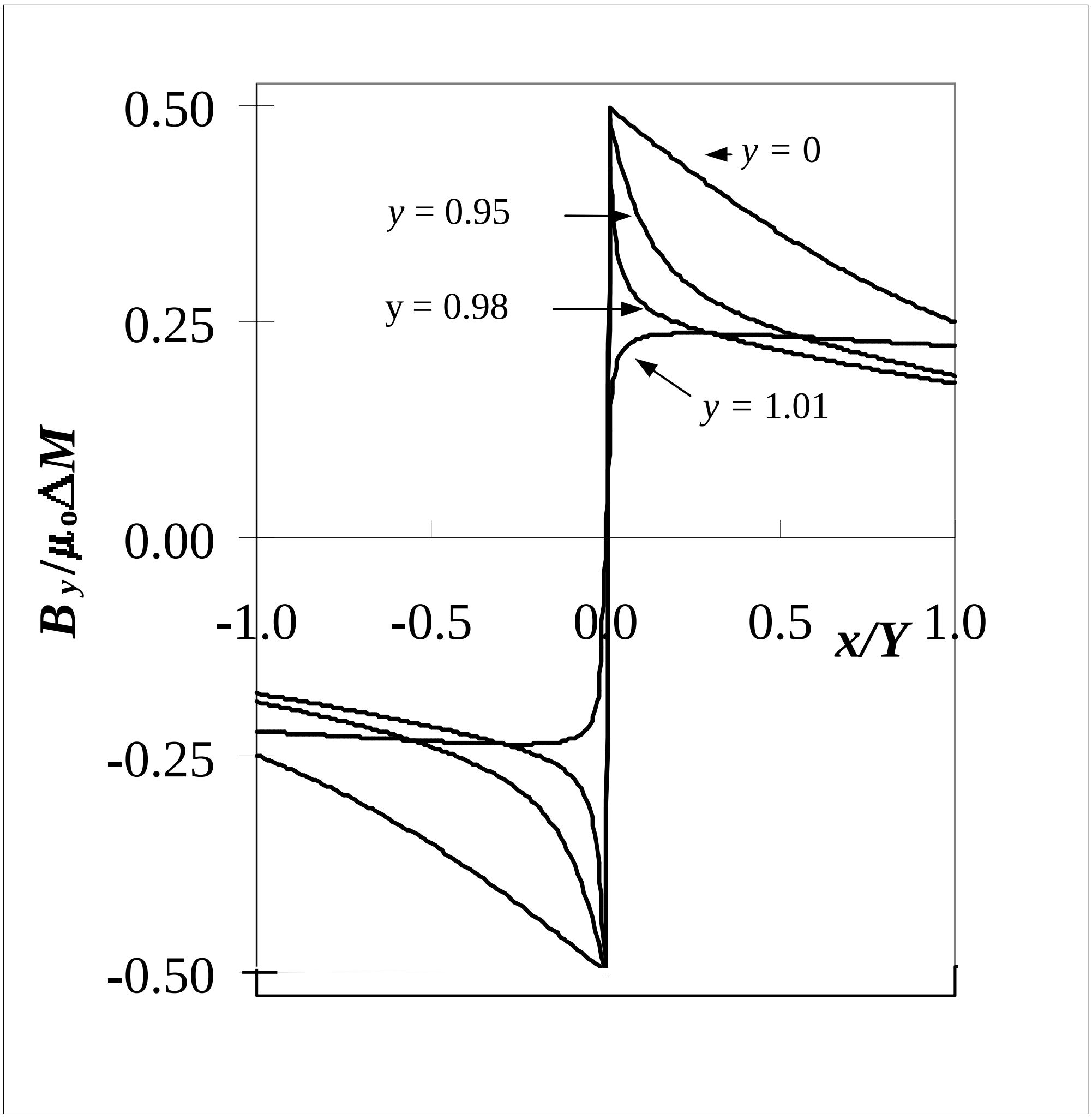

Figure 2 The *B* field due to a current sheet in the $x = 0$ plane of a long slab. Shown are fields calculated as a function of distance along the slab, at a number of values of the distance from the centre line, $y = 0$, expressed as fractions of the slab half-thickness.

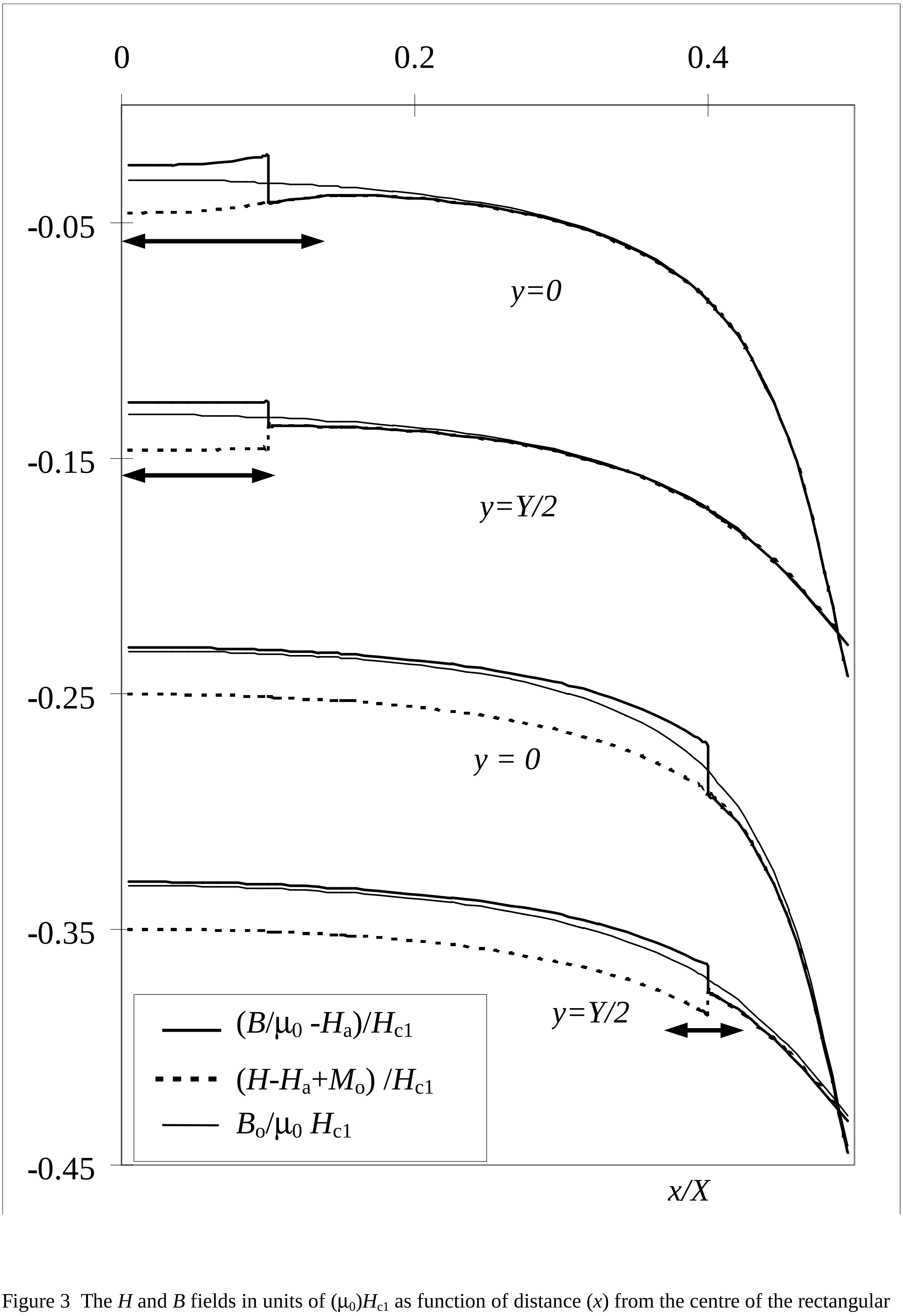

Figure 3 The *H* and *B* fields in units of $(\mu_0)H_{c1}$ as function of distance (*x*) from the centre of the rectangular specimen illustrated in Figure 1(c), assuming a current sheet near the centre (in the *x* direction) of the sample and near the edge. Variations along the line $y = 0$ and along the surface of the sample are shown in each case. The predicted range of the intermediate state is indicated by the arrows. In all cases $\Delta M$ is assumed to be 0.02 $H_{c1}$. $H_0$, $B_0$ and $M_0$ have been calculated assuming no phase transition and therefore represent the 'background' field variations. The different plots have been separated vertically by 0.1 $H_{c1}$ in the interests of clarity.

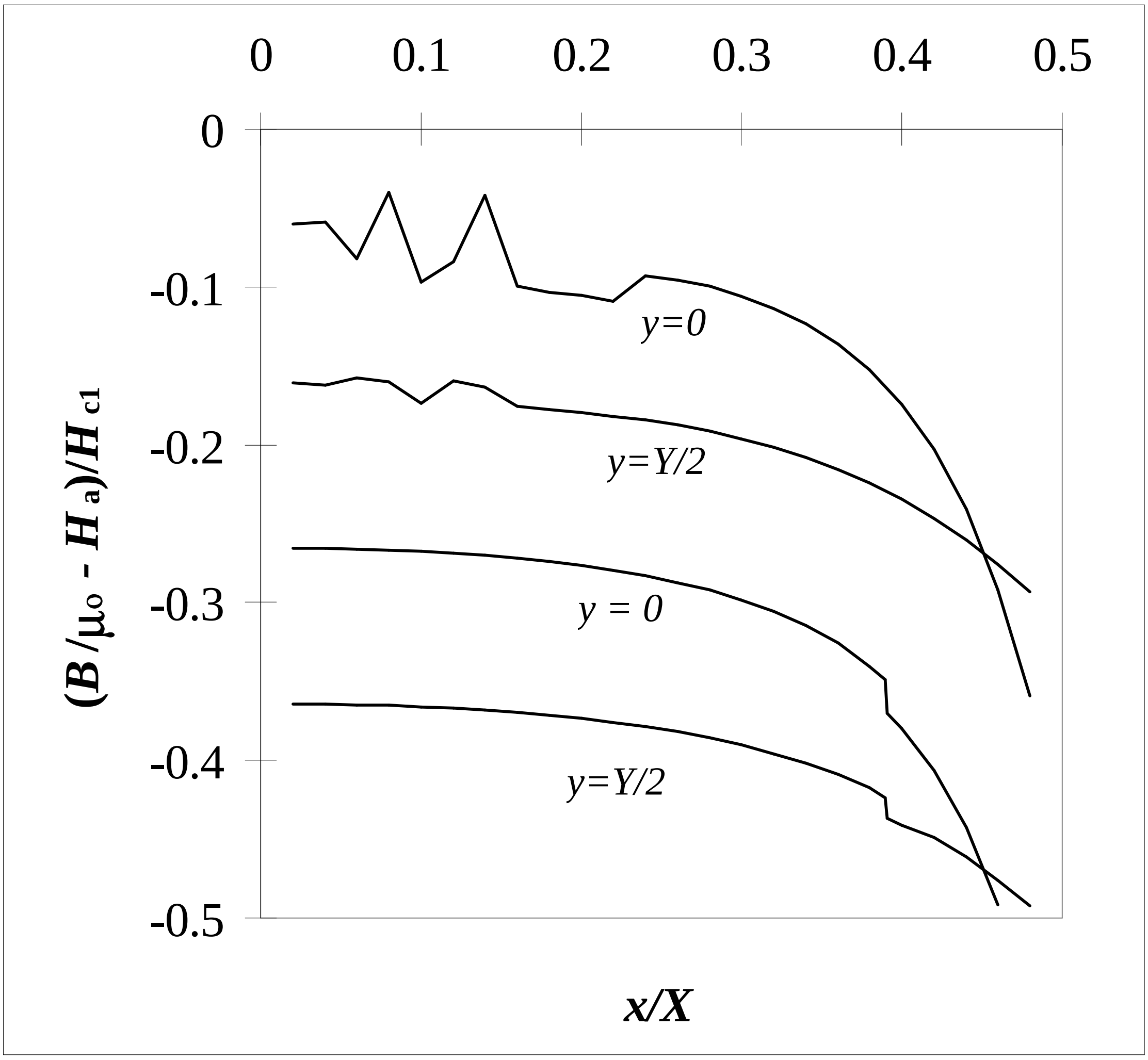

Figure 4a

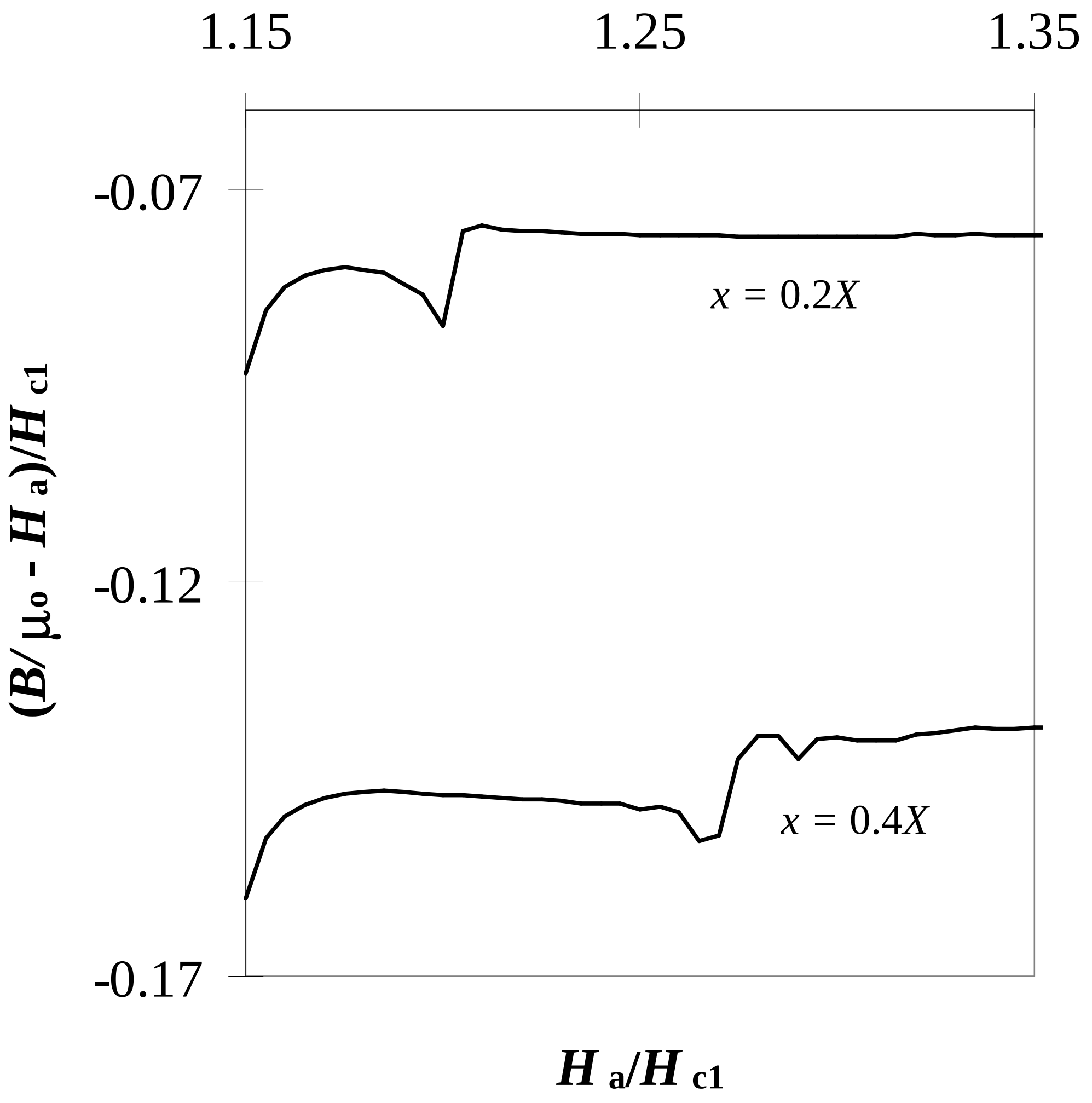

Figure 4b

Figure 4 (a) The variation of the *B* field for a rectangular specimen, as function of distance ($x$) from the centre of the rectangular specimen illustrated in Figure 1(c), calculated numerically in the manner described in the text. Variations along the line $y = 0$ and along the surface of the sample are shown in each case. The applied field is $1.19H_{c1}$ in the case of the upper two plots and $1.22\ H_{c1}$ for the lower pair. The different plots have been separated vertically by $0.1\ H_{c1}$ in the interests of clarity. Similar data are presented as functions of applied field in (b).

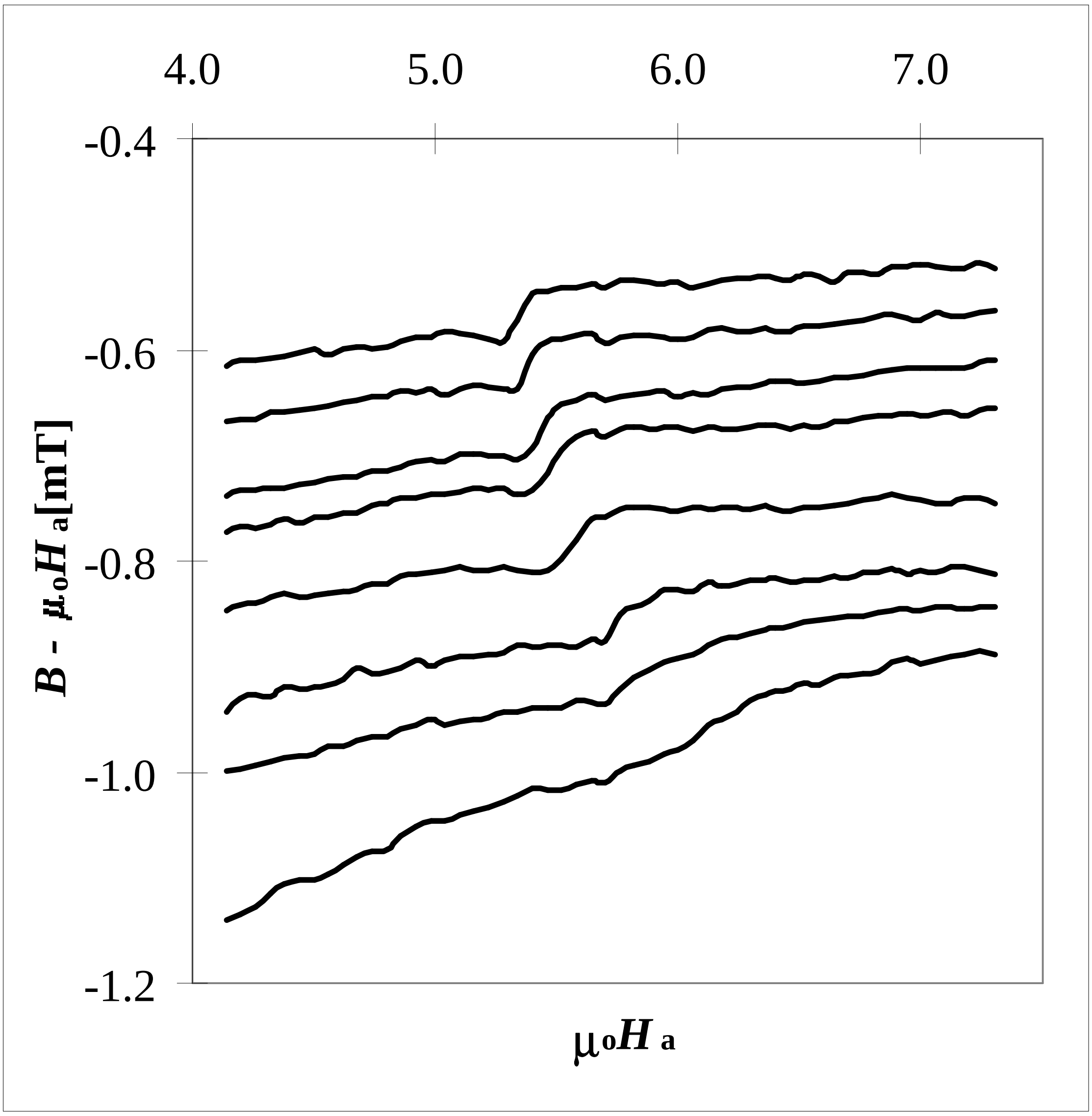

Figure 5b

Figure 5 (a) Comparison of local magnetisation, measured using Hall probes, with global magnetisation using a SQUID magnetometer, for the same crystal. The data was measured at 80K in both cases. The dashed lines are extrapolations of the slope of the behaviour above and below melting. The size of the melting step is clearly about twice as large in the global case compared with the local measurements.
(b) Experimentally measured values of magnetic induction for a rectangular specimen. The lowest trace corresponds to a sensor very near the edge of the crystal and the others are successively nearer the centre with the top one about half way in. Data for increasing and decreasing fields have been averaged.

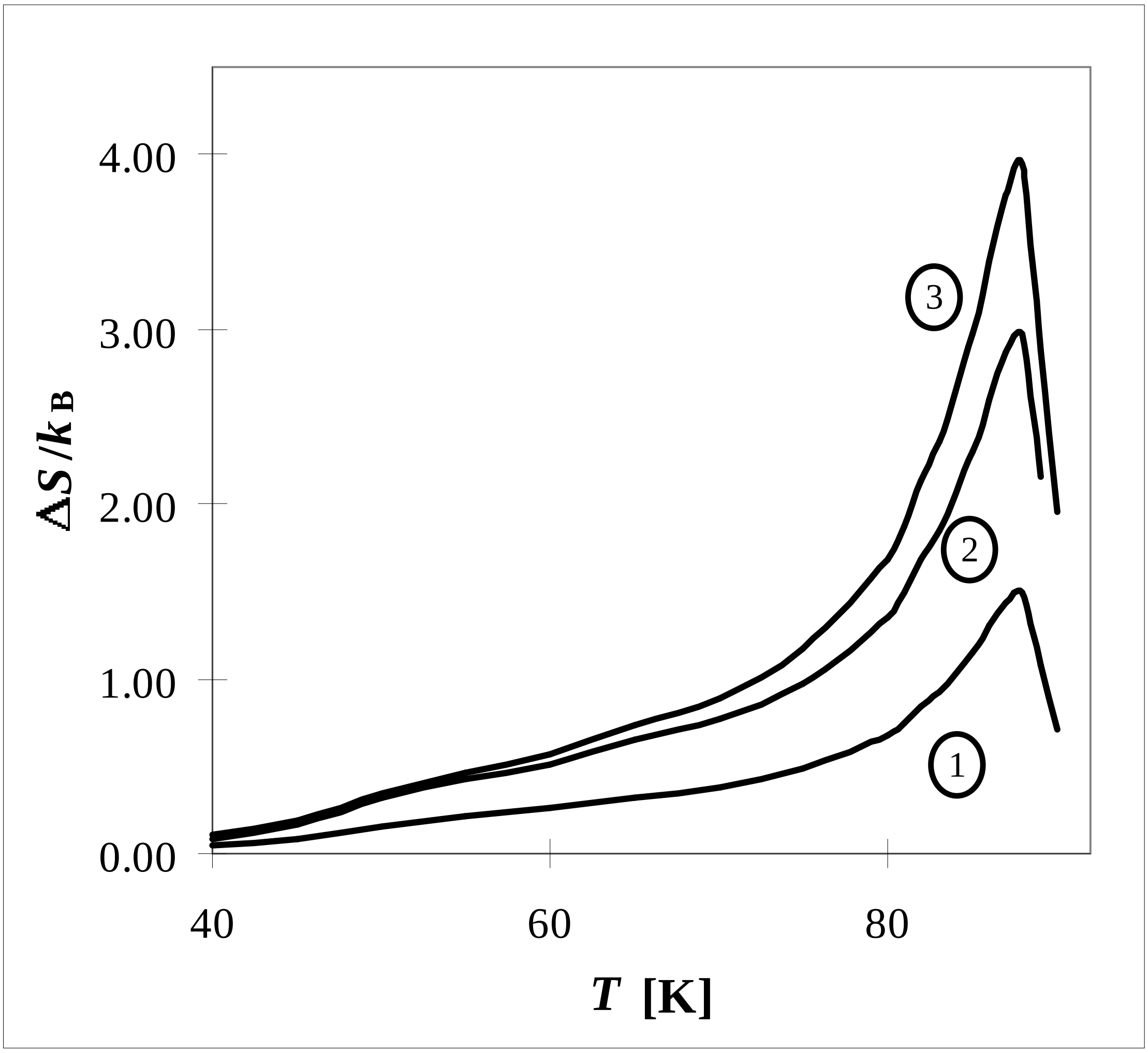

Figure 6 The entropy change at the phase transition in units of $k_B$ per flux line per layer as a function of temperature, calculated from the local field data reported [1]. Graph (1) is calculated as in [1], graph (2) includes the factor of two relating the field change at the surface to the magnetisation, and graph (3) is as (2), but using $dH_m/dT_m$ instead of $dB_m/dT_m$ in the Clausius-Clapeyron equation.

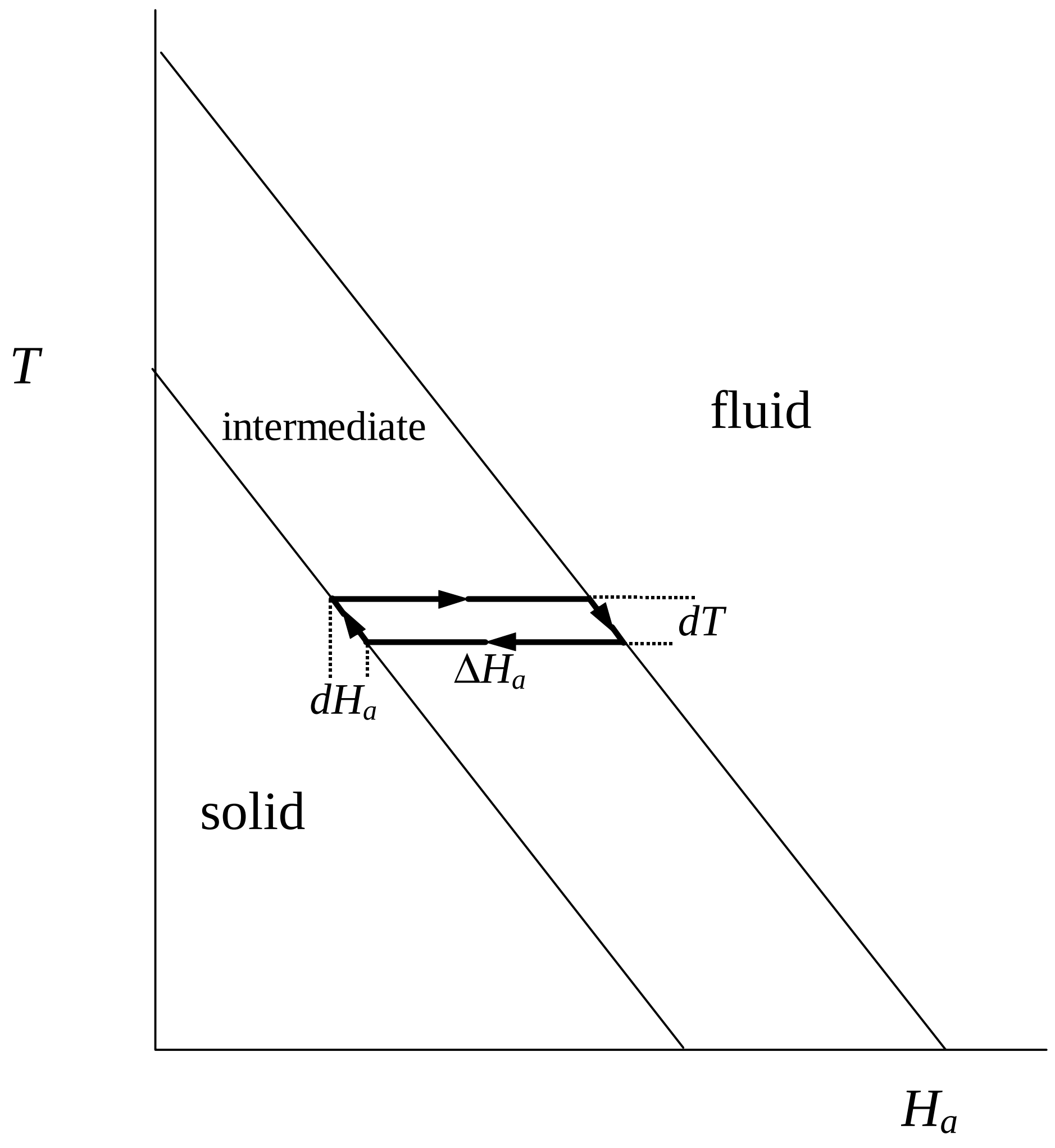

Figure 7  The application of the Clausius-Clapeyron equation to the case of a phase transition which involves an intermediate state.

**The Interpretation of Magnetisation and Entropy Jumps in the Mixed State of High-Temperature Superconductors.**

A.I.M. Rae and E.M. Forgan,
Department of Physics and Space Research, University of Birmingham,
Edgbaston, Birmingham B15 2TT, U.K.

R. A. Doyle
IRC in Superconductivity, University of Cambridge, West Cambridge Site,
Madingley Road, CB3 0HE, U.K.

**Abstract**

In the high-temperature superconductor BSCCO, local measurements of magnetic field at the surface of a crystal in the mixed state show sharp changes as a function of applied field or temperature. These 'jumps' have been interpreted as signs of a first-order flux lattice melting (or sublimation) transition. We show that if 'intermediate state' effects are accounted for, a first-order transition leads to a sharp jump in the global magnetisation only in the case of samples that are significantly non-ellipsoidal in shape. We also investigate the relationship between a jump in magnetisation, *M*, and the associated change in the *B*-field immediately above the crystal surface and show that $\Delta M$ is expected to be twice $\Delta B/\mu_0$. In addition, we emphasise that the Clausius-Clapeyron relationship between magnetisation jump and entropy jump should involve the local *H*-field, not the *B*-field or the applied *H*-field. Re-interpreting some published experimental data taking these factors into account leads to the conclusion that the entropy change can be as much as 4.0 $k_B$ per flux line per layer, compared with less than 2 $k_B$ previously reported for this data, and that a similar factor should be applied to other measurements, where the entropy change can now be as high as 14 $k_B$ per flux line per layer. We show that part of this entropy can be attributed to the cores of the extra flux lines introduced into the sample by the transition and that a considerable amount of the remainder may be associated with changes in the microscopic degrees of freedom

We present and analyse new experimental data on local field jumps and global magnetisation measurements and find that they agree with the above. We also show that these data are consistent with the boundary region between the liquid and solid phases having a width of around 20 flux-line spacings at a field of 10 mT.

## Introduction

The first strong thermodynamic evidence suggesting that flux lattice melting in HTC's might be a first order phase transition was obtained by local measurements of magnetic field (using micro Hall bars) on a sample of BSCCO [1,2]. Earlier non-thermodynamic evidence included sharp discontinuities in the resistivity of YBCO first seen by Safar et al. [3]. Since that time, somewhat broader magnetisation jumps in BSCCO [4] and YBCO [5] have been observed by macroscopic magnetic measurements and recently the entropy jumps that must accompany any true thermodynamic magnetisation jumps have been directly observed in YBCO by calorimetric measurements; good agreement between the two estimates has been obtained in one case[6], although not in another [7]. However, in BSCCO no direct calorimetric measurements of entropy jumps have yet been reported. The interpretation of all the observed jumps in terms of a true thermodynamic phase transition remains controversial [8]. In addition, as was first pointed out by Zeldov et al. [1], it is difficult to reconcile the apparent magnitude of the entropy jump which these workers reported to be up to about $2k_B$ per flux line per layer, with the predictions of any simple melting theory that involves only the degrees of freedom associated with disordering the vortices. Indeed, later measurements on another crystal of BSCCO [9] indicate even larger entropy jumps of up to $6k_B$ per flux line per layer. Recent attempts to understand this have mostly concentrated on YBCO and involved Monte Carlo or Langevin studies including the possible effects of vortex loop unbinding models [10] and separate melting and de-coupling (or other) scenarios [11,12]. Hu and MacDonald [13] have recently carried out Monte Carlo simulations in YBCO and they suggest that the largest fraction of the entropy comes from changes on microscopic length scales and not the entropy content of the vortex configurations, although their analysis probably does not extend to very anisotropic materials like BSCCO. Dodgson et al. in another preprint [14] apply a similar reasoning to the London model and extend this to BSCCO.

As we shall show, the relationship between the phase transition and local magnetic field jumps in rectangular samples is non-trivial Indeed, in an ellipsoidal sample, discontinuous local field jumps may not be expected. This is because a transition involving a sharp positive jump in $M$ at a particular value of $H$ cannot occur in an ellipsoidal specimen, for the same reason as the S-N transition of a type-I superconductor in a magnetic field does not undergo a sharp jump in magnetisation at a fixed value of applied field, but passes steadily from the superconducting to the normal state via the intermediate state. Furthermore, the correct form of the Clausius-Clapeyron equation relating magnetisation jumps to entropy jumps involves the change with temperature of the local H-field (which is not measured by the local sensor), rather than the local B-field (which is). This means that the entropy jumps associated with the phase transition are even larger than previously estimated, especially near $T_c$.

The overall aim of this paper is to re-analyse typical experimental situations in terms of the basic equations of classical macroscopic electromagnetism and thermodynamics and to re-interpret the experimental data to provide improved estimates of the changes in magnetisation and entropy at the phase transition in the flux-line system. We report previously unpublished experimental data and analyse it to obtain values for the width of the interface between the two phases in a BSCCO crystal

**Analysis of Magnetic Properties**

We derive the expected change in the normal component of the $B$ field at the surface of a rectangular specimen of superconductor. We assume that a phase change takes place at a particular value, $H_m$, of the local $H$ field and that this transition is accompanied by a positive jump $\Delta M$ in the magnetisation, which has the (negative) value $M_m$ just below the transition. The 'fluid' state with larger (i.e. less negative) magnetisation corresponds to $H > H_m$ and the 'solid' state occurs when $H < H_m$.

We first consider the case of an ellipsoidal specimen with a demagnetising factor, $\gamma$, (which is zero for fields parallel to a thin rod and unity for fields perpendicular to a zero-thickness plate) and subject to an applied field, $H_a$. The following relations hold [15]:

$$H = H_a - \gamma M$$
$$B = \mu_0 (H + M) = \mu_0 \left[ H_a + (1 - \gamma) M \right] \tag{1}$$

These equations cannot be solved to yield a uniform value of $M$ at all values of $H_a$ if there is a positive jump in $M$ at a given $H = H_m$. This is closely analogous to the transition at $H_c$ of a type I superconductor. Just as in this case, we expect an intermediate state to be set up where the specimen is split into regions of solid and fluid (superconductor and normal for type I). The thickness of these is determined partly by the interface energy (see below), but if this is small they will be thin compared to the thickness of the specimen - see Figure 1a. It follows from this and the above equations that the fraction of specimen in the fluid state increases from zero when $H_a = H_m + \gamma M_m$, and fills the specimen when $H_a = H_m + \gamma(M_m + \Delta M)$. The average magnetisation therefore increases linearly from $M_m$ to $M_m + \Delta M$ over this range in $H_a$, as shown in Figure 1b. We note that the change in $(B/\mu_0\text{-}H_a)$ is much smaller than $\Delta M$ when $\gamma \sim 1$.

*Rectangular Specimens*

If the sample shape is non-ellipsoidal, the *H*-field inside the specimen is non-uniform and the physics cannot be described by a simple demagnetising factor, because the 'background' field in the absence of a transition now varies across the specimen.

We limit ourselves to 'slab-like' crystals whose thickness in the direction of applied field is always smaller much than the other crystal dimensions (although large compared with the superconducting penetration depth). In this case the 'background' field varies across the width of the specimen and the thicker it is, the greater is the field gradient. Also the field gradient increases with distance from the centre of the crystal where it is zero. We might therefore expect to see a magnetic field jump at a particular position in the crystal where the local *H*-field is equal to the value at which the phase transition occurs. However, for sufficiently small field gradients a finite region of intermediate state may be expected. We can make this argument more quantitative by making some reasonably realistic simplifying assumptions:

1. that the specimen is a flat plate which is infinitely long in one direction, so that we can treat the problem as 2-dimensional

2. that the magnetisation has a value $M_0$ which is independent of the local $H$ field (except for the small positive jump ($\Delta M$) due to the phase transition). This is a good approximation for a high $\kappa$ superconductor, provided $H$ is appreciably greater than $H_{c1}$ [16]. Typical observed $H_m$'s for BSCCO are between 1.5 and $2.0H_{c1}$ with $M_0 \sim -0.5H_{c1}$ in this region.

3. that the magnetisation is everywhere parallel to the c axis, which is quite accurate for a very anisotropic superconductor such as BSCCO.

4. $\Delta M = 0.02H_{c1}$; this value was chosen in the interests of clear presentation and is about twice the maximum measured in BSCCO.

Referring to Figure 1c, it follows from 2 and 3 above that the overall magnetisation is equivalent to current sheets at the edges of the crystal and the 'background' variation in field through the specimen is due to the field from these current sheets. In addition, we postulate that the fluid region is confined to a central region of the specimen with the outer parts solid. The magnetisation jump associated with the transition is then equivalent to current sheets on the planes dividing the solid and liquid regions (see Figure 1c again). An analytic expression for the *B*-field as a function of position is then readily obtained from the standard expression for the field due to a current sheet separating two regions where the magnetisations differ by $\Delta M$

$$B_y = \frac{\mu_0 \Delta M}{2\pi}\left[\tan^{-1}\left(\frac{Y/2 - y}{x}\right) + \tan^{-1}\left(\frac{Y/2 + y}{x}\right)\right] + B_0 \qquad (2)$$

where $y$ is parallel to the current sheet (i.e. perpendicular to the crystal surface) and $x$ is the perpendicular distance from the sheet. The origin in the y direction is at the centre of the current sheet which has a width $Y$. $B_0$ is the 'background' field that would exist in the absence of this current sheet.

Figure 2 shows a plot of this function (with $B_0 = 0$) in the vicinity of the current sheet and for various values of $y$ (as a fraction of $Y/2$). We note that the *B*-field at the $y = 0$ mid-line jumps by $\mu_0\Delta M$ as the current sheet is crossed. At higher values of $y$, the jump at $x = 0$ is still $\mu_0\Delta M$, but $B_y$ quickly drops to about half this value in a narrow region in $x$ of width ~ $(Y/2 - y)$. Just below this top surface, if the immediate vicinity of the current sheet is ignored, the jump appears to be $\mu_0\Delta M/2$, and this is accompanied by a jump in $H$ below the surface of $\Delta M/2$ *in the opposite direction*. This apparently breaches the standard boundary conditions which imply that $H_y$ should be continuous, but the narrow spike in $B$ in the immediate vicinity of the sheet restores the expected continuity. Just above the top surface, $B = \mu_0 H$, the jump in both quantities is $\mu_0\Delta M/2$, and there is no singularity as the current sheet is crossed.

The factor of two between the jump in $B$ at the surface and that in the centre of the crystal can be understood on symmetry grounds. The value at the mid point of a current sheet - $\mu_0\Delta M$ - is made up from identical contributions from the two halves of the current sheet that lie above and below the $y = 0$ plane. If we now imagine the specimen sliced along this plane and remove one half of it, the plane becomes a surface and $\Delta B$ is halved. (A similar argument is often employed to show that the field at the end of a long solenoid is half that far from the end.) We also note that the sharp spike in $B$ is associated with the divergence of the argument of the first term in ( 2); it follows that this effect will be largely independent of any curvature of the current sheet unless this is very close to the surface.

Figure 3 shows the $B$-field and $H$-field as a function of position in a long crystal for which the cross section has an aspect ratio ($Y/X$) of 10 (comparable with that used in many experiments). Current sheets are placed as discussed above, with those representing the melting transition located (a) close to the centre and (b) about 10% from the edge in the $x$ direction (c.f. Figure 1c). The $y$ components of the fields at the surface of the crystal and along the line $y = 0$ (c.f. Figure 1c) are shown in each case. We note the following points:

1. As expected, the $B$-field at the $y = 0$ mid-line jumps by $\mu_0\Delta M$ as the current sheet is crossed. When this jump is near the edge of the crystal (in the $x$ direction), the change in the $H$-field is monotonic and no intermediate state should arise. Near the centre of the crystal, however, there is a region on either side of the current sheet where the $H$-field varies in the 'wrong way' (i.e. it is low in the high-field phase and high in the low-field phase which is not consistent with the assumed $M$-$H$ relation) and therefore where an intermediate state could exist.

2. At the surface of the crystal the discontinuity in $B$ is indeed $\mu_0\Delta M/2$ and there is a similar, but opposite, jump in $H$. This means that, within the limits of this model, there is always a region over which an intermediate state should be expected. As is shown in Figure 3 this region is quite large near the centre where the 'background' field gradient is low, but near the edge of the crystal it is much smaller and could be comparable to the size of a typical detector (~ 5μm or about 0.1x the crystal thickness).

3. Estimates of the entropy jump at the transition based on local field measurements at the crystal surface and assuming that $\Delta B = \mu_0\Delta M$ are underestimated by a factor of two. This means that the maximum entropy jump observed in [1] is $3.0k_B$ per pancake, while that reported in [9] should actually be about 12 $k_B$.

The above calculation assumes that the current sheet was at some arbitrary fixed position in the crystal rather than being defined by the locus of points along which the *H* field is equal to the melting field. We have therefore attempted a more realistic simulation based on a numerical calculation, using a modification of the standard relaxation method [17]. We represent the *H* field as the gradient of a potential which we evaluate on a two-dimensional grid of points, spanning the rectangular specimen and a surrounding area 100 times larger than that of the specimen. As in the analytical calculation described above, we assume that the magnetisation is always in the *y* direction and uniform over the area of a grid unit surrounding a grid point and that the specimen has an aspect ratio of 10. Consistent with the simplifying assumptions set out above, *M* is taken to be a function of the *y* component of the *H*-field at the grid point given by

$$\begin{aligned} M\left(H_y\right) &= -H_y, \text{ if } H_y \le H_{c1} \\ &= -0.5 * H_{c1} \text{ if } H_{c1} < H_y \le H_m \\ &= -0.5 * H_{c1} + \Delta M \text{ if } H_y > H_m \end{aligned} \tag{3}$$

Using a second-order Taylor expansion of the potential about a grid point along with the condition that the normal component of the *B* field is always continuous, we obtain the following expression for the potential at a grid point in terms of the potential and the magnetisation at surrounding grid points

$$\phi_{ij} = \frac{\left(\phi_{i+1,j} + \phi_{i-1,j}\right)\delta y^2 + \left(\phi_{i,j+1} + \phi_{i,j-1}\right)\delta x^2 + \left(M_{i,j+1} - M_{i,j-1}\right)\delta x^2 \delta y}{2\left(\delta x^2 + \delta y^2\right)} \tag{4}$$

where $\phi_{ij}$ is the potential at the grid point whose co-ordinates are $(i\delta x, j\delta y)$ We can also calculate the components of the *H*-field as the gradient of the potential to get

$$H_x = \frac{\phi_{i+1,j} - \phi_{i-1,j}}{2\delta x}$$

$$H_y = \frac{\phi_{i,j+1} - \phi_{i,j-1}}{2\delta y} + \frac{M_{i,j+1} + M_{i,j-1} - 2M_{ij}}{4} \qquad (5)$$

Boundary conditions are set at the edges of the grid so that the field is equal to the applied field, $H_a$, far from the specimen.

Equations ( 4) and ( 5) are iterated, looking for self consistency, using a grid size similar to the separation between local probes in a typical experiment on typical sample. When the applied field is such that $H > H_{c1}$ everywhere in the specimen, but is not equal to the melting field $H_m$ anywhere, self consistency is readily achieved and the *H*-field is essentially identical to the analytic form set out in ( 2) above - assuming current sheets at the outer edges of the specimen. When $H_a$ is chosen so that $H = H_m$ close to the edge of the specimen, convergence is again quite rapid and a distinct jump is observed in the *B*-field at the surface of the specimen as well as at the middle (in the *y* direction). Moreover, the sizes of the jumps are close to those predicted earlier with the centre one being twice that at the surface. In the case of smaller $H_a$ where $H = H_m$ near the centre (in the *x* direction), convergence is not achieved. This is because there is no self-consistent solution to the equations for the reasons given earlier and the system generally oscillates between two or more states. The *B*-field in such a case is shown in Figure 4a. We note that it oscillates as a function of distance along the specimen in a manner reminiscent of the form of the intermediate state shown in Figure 1a. The amplitude of the oscillations along the centre line is about twice that at the surface, and the region over which this structure exists is similar to that predicted from the analytical calculations shown in Figure 3. Similar instabilities have been observed by other workers, where the M(H) behaviour of BSCCO disks has been modelled and the effects of melting investigated [18]. However, all these results have to be treated with some caution because there is no convergence of the iterative process in the regions where self consistency cannot be achieved. Experimentally, no direct evidence has yet been found for an intermediate state although structure in the field profiles in the vicinity of melting has recently been reported and ascribed to a possible disorder-related intermediate state [19].

It appears from Figure 4a that one difference between the numerical results and the analytical ones is that the horizontal position of the transition at the centre of the slab (in the *y* direction) is nearer the edge of the crystal (in the *x* direction) than is the transition at the surface. This is consistent with the form of the *y* dependence of the background field, as observed in a simulation carried out with similar sized fields applied to the same sized specimen, but with no phase transition included. This implies that the current sheet is not always vertical as was assumed in the analytic calculation, although it should be noted that the effect is most pronounced in the region where convergence of the iteration was not fully achieved. Nevertheless, the jump in the *B*-field at the centre is still equal to $\mu_0\Delta M$ and that at the surface is still half this value. The main conclusions following from the analytic calculation therefore still stand. However, we should note that the assumptions underlying all our calculations include a zero-thickness interface and macroscopic fields, and these can be expected to be less valid at length scales comparable with the flux-line spacing.

In contrast with the strong suggestions from both theoretical approaches above, typical experimental signals from Hall probes at the surface and close to the centre (in the *x* direction) of the sample show very sharp jumps [1]. This is partly due to data being plotted as functions of temperature or applied field rather than position in the specimen: given that the background field gradient is very small at the centre, the phase boundary will cross the crystal very rapidly as one of these quantities is changed. This is exemplified in Figure 4b which shows the numerical simulation data plotted as a function of applied field. The transitions now appear appreciably sharper, although there is still significant structure over an applied field range amounting to a few percent of $H_{c1}$, and therefore indicating an intermediate state which has not been reported experimentally. One reason why the intermediate state could be suppressed completely in real specimens would be the presence of a positive surface energy associated with the phase boundary. If this were large enough, there would be an energy cost from the formation of the additional surfaces associated with intermediate state. We can estimate the magnitude of the surface energy required to do this. A simple thermodynamic argument leads to the conclusion that the free energy per unit volume of material in the 'wrong' phase is $\mu_0 H_m \Delta M$. The surface energy that would be required to make a sharp boundary preferred is therefore equal to $\mu_0 H_m \Delta M d$, where *d* is the distance over which the intermediate state would form in the absence of such a surface energy.

*Experimental Investigations*

Isothermal magnetisation measurements were made on a crystal of dimensions 300x240x40 $\mu m^3$ grown by the floating zone technique [20], using a Quantum Design MPMS5 SQUID magnetometer with due care taken to reduce the remanent field. A 3cm scan length was used to minimise the effects of moving the sample through the weakly inhomogeneous field. The results of such a measurement performed at a temperature of 80K are shown in Figure 5a.

Local magnetisation measurements were made on the same crystal using a miniature array of GaAs/AlGaAs two dimensional electron gas (2DEG) Hall sensors which have both high sensitivity and high spatial resolution. The active area of the sensors was 2.5 x 5 $\mu m^2$ and a linear array of 9 sensors each spaced (centre to centre) by 5 μm was used. The sample is optically smooth. The series of nine detectors was placed directly onto the crystal surface perpendicular to the shortest dimension: one sensor was positioned very close to, but outside of, the edge of the crystal and the others were placed in a row perpendicular to this edge, extending about half way to the crystal centre. The response, $B_z$, can be measured as a function of applied field or temperature. The 'local magnetisation' is defined as the difference between the measured $B_z$ and the applied field, $\mu_0 H_a$. The results of one of these experiments is included in Figure 5a, where the signals recorded as the applied field was changed at a constant temperature of 80K. Figure 5b shows similar data from each of the eight detectors that are within the specimen area.

A direct test of the predicted halving of the jump in the surface field compared to the change in magnetisation is to compare the local field measurements with the global magnetisation measurements for the same specimen under the same conditions. It is clear from Figure 5a that the magnetisation change is indeed twice the jump in the local field within experimental error. We are aware of only one previous attempt to do this [21] where $\mu_0 \Delta M$ was measured to be about 5 times $\Delta B$. This larger than expected factor could result if the local field detectors were a small, but significant, distance above the surface.

We see that the transition measured from the local field data (Figure 5b) is quite sharp for the detector near the centre, but becomes considerably broader as the edge of the sample is approached. Moreover, it is also clear that the separation between the mid points of the transitions in neighbouring traces is also increasing. We can calculate the gradient of the $H$ field in the specimen from the latter data by assuming that the mid point of the transition is always at the same value of the local $H$-field. We note that the results of this procedure are consistent with those calculated by the methods discussed earlier for a similarly shaped crystal. We can now use the field gradients to convert the widths in $H_a$ into widths in position and find that the width in $x$ appears to have no systematic dependence on the detector position and has an average value of 2.3+-0.3 detector separations. Allowing for the finite resolution of the detector, this result is consistent with a phase-boundary width of about 1.5 detector spacings, which is 7.5μm or about 14 flux-line separations at the fields used. It should be noted that if this width were associated with the intermediate-state region, it would be expected to become smaller as we moved away from the centre of the crystal, which is contrary to what we observed.

## Thermodynamic Calculations

In Zeldov et al. [1], an expression is given for the entropy change, $\Delta S_1$, associated with the first-order phase transition. If we convert this to SI units and apply the factor of 2 discussed in the previous section, we obtain

$$\Delta S_1 = -\frac{2}{\mu_0}\frac{dB_m}{dT_m}\Delta B \text{ per unit volume.} \tag{6}$$

$B_m$, and $T_m$ are the values of the magnetic flux density and temperature at the transition, and $\Delta B$ is the discontinuity in the $B$-field at the surface. The resulting values of $\Delta S_1(T)$ are therefore twice those obtained by Zeldov et al. [1]; they are shown (expressed per 'pancake' volume corresponding to the area occupied by a flux line multiplied by the inter-layer separation) in Figure 6 where they are seen to rise to a value of about $3k_B$ per pancake close to $T_c$. This is considerably larger than would be expected from a simple disordering of the pancakes, as has been confirmed by Monte-Carlo simulations [12,13] which predict $\Delta S \sim 0.5k_B$ per pancake or less.

The standard expression for the entropy based on the Clausius-Clapeyron relation at a first-order magnetic transition [15] is

$$\Delta S_2 = -\frac{dH_m}{dT_m}\mu_0 \Delta M$$

( 7)

which differs from ( 6) in two respects. First, $\mu_0\Delta M$ replaces $\Delta B$, which introduces the factor of 2 discussed above. Secondly, $dB_m/dT_m$ is replaced by $\mu_0 dH_m/dT_m$. From elementary electromagnetism, the difference between these expressions is $\mu_0 dM_m/dT_m$. There are a number of expressions for the magnetisation of a type II superconductor depending on the value of the applied field relative to $H_{c1}$ and $H_{c2}$, but in all cases the temperature dependence is proportional to that of $H_{c1}$ apart from possible logarithmic terms which we ignore. In the absence of direct measurements of this quantity, we therefore put

$$\frac{dM_m}{dT} = -f\,\frac{dH_{c1}}{dT}$$

( 8)

where $f$ is a numerical constant whose value can be deduced from the experimental magnetisation curves [1,2] as ~ 0.5 . Assuming that the temperature dependence of $H_{c1}$ is given by

$$H_{c1}(T) = H_{c10}\left(1 - \frac{T^n}{T_c^n}\right)$$

( 9)

and using the standard relations for a type II superconductor [16], we get

$$\frac{dH_m}{dT_m} = \frac{1}{\mu_0}\frac{dB_m}{dT_m} - f\,\frac{nT_m^{n-1}}{T_c^n}H_{c10} = \frac{1}{\mu_0}\frac{dB_m}{dT_m} - f\,\frac{n\phi_0 T_m^{n-1}}{4\pi\lambda_0^2 T_c^n}\ln\kappa$$

( 10)

We have calculated $\frac{dH_m}{dT_m}$ in the case of BSCCO, using the data of Zeldov et al. for $\frac{dB_m}{dT_m}$ and assuming $T_c$ = 90.9K with $\kappa = 47$, $\lambda_0$ = 210nm and $n$ = 2; the latter two values lead to a reasonable fit to the experimentally measured $\lambda(T)$ [22] over the relevant temperature range. We find that the contribution from the temperature dependence of the magnetisation is significant, contributing up to about 25% of the total near $T_c$. Using equation ( 9) and again assuming that $\mu_0\Delta M$ is twice the jump in $B$ observed at the sample surface, we re-calculate $\Delta S$ from the data of Zeldov et al. [1] and the results are shown in Figure 6. The entropy change is now considerably more than twice that calculated by these authors, reaching a maximum of about $4k_B$ per flux-line per layer at $T$ = 86K. A similar treatment of the data reported by Morozov et al. [2] leads to $14k_B$ per flux-line per layer at $T$ = 89K. These are in even greater disagreement with models based purely on vortex disorder.

It might be thought that $H_m$ should be replaced by the corresponding applied field, as that is the quantity conjugate to $M$ in the case of an ellipsoidal specimen with a finite de-magnetising factor [15]. However, we can show that ( 7) actually applies in this case also because of the finite width of the transition as discussed earlier. Figure 7 shows a closed path about a transition whose width in applied field is $\Delta H_a = \gamma\Delta M$. Taking the free energy to be

$$G = E - TS - \mu_0 M H_a$$

( 11)

so that

$$\left(\frac{\partial G}{\partial T}\right)_M = -S \;\text{ and }\; \left(\frac{\partial G}{\partial H_a}\right)_T = -\mu_0 M$$

( 12)

we put the total change in $G$ on going round the closed path shown in Figure 7 equal to zero to get

$$-S_s dT - \mu_0 M_s dH_a - \mu_0 M_m(T+dT)\Delta H_a + S_f dT + \mu_0 M_f dH_a + \mu_0 M_m(T)\Delta H_a = 0$$

( 13)

where the subscripts $s$ and $f$ refer to the solid and fluid phases respectively and $M_m(T)$ is the average of the solid and fluid magnetisations. (It should be noted that we have started at the bottom left-hand corner of the closed path in Figure 7and that, for the slope shown, $dH_a$ is negative). It follows that

$$\begin{aligned}\Delta S = S_f - S_s &= -\mu_0\left(\frac{dH_{am}}{dT_m}\Delta M - \frac{dM_m}{dT_m}\Delta H_a\right)\\ &= -\mu_0 \Delta M \frac{d}{dT_m}\left(H_{am} - \gamma M_m\right)\\ &= -\mu_0 \Delta M \frac{dH_m}{dT_m}\end{aligned}$$

( 14)

We therefore conclude that ( 7) is universal - i.e. independent of geometry - provided $H_m$ is the local value of the actual $H$ field rather than the applied field and that demagnetisation has been taken into account when calculating $\Delta M$ from the experimental data. This justifies using ( 7) to calculate the entropy change from the local field jump; in contrast, the de-magnetising factor would have to be explicitly included if the data were obtained from a measurement of the average ($B$ - $H_a$) on an ellipsoidal crystal (c.f. Figure 1). On the other hand, a direct measurement of the global magnetisation change, such as that reported earlier, requires no such correction. These statements should clarify any possible ambiguity in [8].

The result in ( 14) is consistent with the fact that the entropy is a function of state, in the sense that it is the integral of a local thermodynamic variable whose value depends on the temperature and the local value of *H*, and which is zero outside the sample. This is in contrast with the free energy whose value depends on those of the magnetic field variables over all space. Provided the corrections we have outlined so far are made to experimental magnetic data, the entropy change will be the same as that which should be observed in the case where demagnetising effects are negligible with the *H*-field is uniform across the specimen and equal to the applied field; this is equivalent to performing experiments in solenoidal geometry.

*The physical significance of the entropy change*

We now consider the question of why the observed entropy change is so much larger than that expected from a simple disordering of the vortices. Typical numerical calculations treat the system as a set of point objects (e.g. pancake vortices) interacting via a temperature-independent potential. However, the standard expressions [16] for the energy of a flux-line lattice are based on Ginzburg-Landau theory, implying that the potential is actually a Helmholtz free energy which includes an entropic term. If the energy were indeed purely internal, the entropy change at the transition would be entirely that associated with disorder in the degrees of freedom describing the positions and motion of the vortices, and there would be no entropy change associated with the changes in the degrees of freedom of the superfluid itself at the transition. However, as first pointed by Hu and MacDonald [13], if the interactions are temperature dependent, the total entropy change will include an additional entropy term associated with the change in what is now part of the Helmholtz free energy.

To understand this further, we write the Gibbs free energies per unit volume of the solid and fluid phases as

$$G_s = F_{s0} - \mu_0 M_s H$$
$$G_f = F_{f0} - \mu_0 M_f H - T\Delta S_m$$

( 15)

where the subscripts $s$ and $f$ refer to the solid and fluid phases respectively. The entropy change associated with disordering of the vortices is denoted as the ‘melting entropy’, $\Delta S_m$. $F_{so}$ and $F_{fo}$ are the remaining contributions to the Helmholtz free energies of the solid and fluid phases; we refer to these as ‘intrinsic’ free energies from now on.

The change, $\Delta F_0$, in the intrinsic (Helmholtz) free energy and the associated change in entropy $\Delta S_0$ are then

$$\Delta F_0 = F_{f0} - F_{s0} \quad ; \quad \Delta S_0 = -\left(\frac{\partial \Delta F_0}{\partial T}\right) \tag{16}$$

As explained above, the transition takes place at constant $T$ and $H$, so at the transition, the Gibbs free energies of the two phases must be equal so, using ( 15) and the first of ( 16), we can express $\Delta F_0$ as

$$\Delta F_0 = T_m \Delta S_m + H_m \Delta B_m \tag{17}$$

where $\Delta B$ is the jump in the B field at the transition and is now equal to $\mu_0 \Delta M$ because we are assuming zero demagnetisation; it is important to note that this equation holds only at the melting line, where $G_s = G_f$, so that $\Delta S_0$ is not equal to the temperature derivative of the right-hand side of ( 17).

One contribution to $\Delta S_0$ is directly associated with the increase in flux density in the fluid phase. We can compare this with the entropy change that would be expected if the same increase in $B$ were to occur as a result of an equilibrium change in the solid phase. Denoting this as $\Delta S_0'$ we have

$$\Delta S_0{}' = \left(\frac{\partial S}{\partial B}\right)_T \Delta B = -\left(\frac{\partial H}{\partial T}\right)_B \Delta B = \left(\frac{\partial M}{\partial T}\right)_B \Delta B$$

( 18)

where we have used a standard Maxwell relation and the fact that the transition occurs at constant *H*.

Physically, $\Delta S_0{}'$ can be described as the additional entropy introduced into the sample when the flux-line density is increased. A large part of this is associated with the free energy of the flux-line cores where the superconducting order parameter is suppressed, but there is a significant contribution from the entropic part of the free energy of interaction between the vortices [23]. In the region of the transition, *M* does not depend strongly on *B* (apart from the jump itself) so $(\partial M/\partial T)_B$ is close to $dM_m / dT_m$. It follows that $\Delta S_0{}'$ is equivalent to our earlier correction which replaced $dB_m/dT_m$ by $dH_m/dT_m$ - see the discussion relating to ( 6) and ( 7) above and Figure 6.

If $\Delta S_0{}'$ were the total change in the intrinsic entropy, this could not explain the experimental data because there is still a strong divergence in the entropy calculated using ( 6) as *T* approaches $T_c$ (Figure 6). Changes in $F_0$ beyond those associated with a simple change in flux-line density are also to be expected as a result of the re-arrangement of the vortices in going from the solid to the liquid phase. We note that if we could assume a temperature dependence for $\Delta F_0$ then we could determine the relation between $\Delta S_0$ and $\Delta S_m$ directly from equations ( 16) to ( 18). This is essentially what has been done by Dodgson et al. [14] in deriving their equation (9): on the assumption that $(\Delta F_0 - H_m \Delta B_m)$ scales directly as $[1 - (T_m/T_c)^2]$, it follows directly that the total entropy change $\Delta S_0 + \Delta S_m$ is equal to $[1 + (T_m/T_c)^2] /[1 - T_m/T_c)^2] \Delta S_m$. This expression may indeed diverge as $T_m$ approaches $T_c$, depending on the detailed temperature dependence of $\Delta S_m$. However, it should be noted that the validity of this scaling is not firmly established, particularly in the case of BSCCO, and that the agreement between experiment and the entropy changes calculated by these authors depends strongly on the assumed form of the melting line.

**Conclusions**

We have shown that the positive jump in magnetisation associated with the flux-line melting transition implies that an ‘intermediate state’ should form over a range of applied fields for ellipsoidal specimens and for rectangular specimens over a volume in the vicinity of the transition line, unless it is suppressed by a positive interface energy. We have shown that jumps in *B* field measured by local surface probes can be reliably interpreted in terms of discontinuities in the magnetisation, provided it is realised that these jumps are half the size of those at the centre of the crystal. We have reported new experimental data that confirms this point directly and also provides evidence for the thickness of the solid-liquid interface. The magnetisation change can in turn be related to the entropy jump associated with the phase transition, provided the correct form of the Clausius-Clapeyron equation is used
Re-interpretation of published experimental data in the light of these considerations leads to the conclusion that the entropy changes associated with the phase transition in BSCCO are more than twice those previously reported. We have confirmed suggestions by other workers that it is reasonable to expect that a considerable amount of the entropy change is associated with the change in the free energy of interaction between the vortices rather than that directly associated with their disorder. Some, though not all, of this can be attributed directly to the reduction in superconducting fraction associated with the increase in flux density in the fluid state.

**Acknowledgements**

We are grateful to Eli Zeldov and Archie Campbell for useful discussions. We thank H Motohira for supplying the BSCCO crystal, RD thanks Dani Majer for advice and assistance with the Hall probe experiments which were carried out in the Weizmann Institute. One of us (AIMR) would also like to thank Bill Powell who many years ago taught him about the importance of understanding **B** and **H**.

**Figure Legends**

Figure 1 An intermediate state consisting of alternate regions of solid and liquid is illustrated in (a) for the case of an ellipsoidal specimen with a field applied along the short axis. The predicted variations of *H* and *B* with applied field $H_a$ as the ellipsoidal specimen passes through the intermediate state are shown in (b). The current-sheet model used to study the behaviour of rectangular specimens is illustrated in (c).

Figure 2 The *B* field due to a current sheet in the $x = 0$ plane of a long slab. Shown are fields calculated as a function of distance along the slab and at a number of values of the distance from the centre line, $y = 0$, expressed as fractions of the slab half-thickness.

Figure 3 The *H* and *B* fields in units of $(\mu_0)H_{c1}$ as function of distance (*x*) from the centre of the rectangular specimen illustrated in Figure 1(c), assuming current sheets at the edges of the sample as well as near the centre (in the *x* direction) or near the edge. Variations along the line $y = 0$ and along the surface of the sample are shown in each case. The predicted range of the intermediate state is indicated by the arrows. In all cases $\Delta M$ is assumed to be 0.02 $H_{c1}$. $H_0$, $B_0$ and $M_0$ **???** have been calculated assuming no phase transition and therefore represent the 'background' field variations. The different plots have been separated vertically by 0.1 $H_{c1}$ in the interests of clarity.

Figure 4 (a) The variation of the *B* field for a rectangular specimen, as function of distance (*x*) from the centre of the rectangular specimen illustrated in Figure 1(c), calculated numerically in the manner described in the text. Variations along the line $y = 0$ and along the surface of the sample are shown in each case. The applied field is $1.19H_{c1}$ in the case of the upper two plots and 1.22 $H_{c1}$ for the lower pair. The different plots have been separated vertically by 0.1 $H_{c1}$ in the interests of clarity. Similar data are presented as functions of applied field in (b).

Figure 5 (a) Comparison of local magnetisation, measured using Hall probes, with global magnetisation using a SQUID magnetometer, for the same crystal The data was measured at 80K in both cases. The dashed lines are extrapolations of the slope of the behaviour above and below melting. The size of the melting step is clearly about twice as large in the global case compared with the local measurements.

(b) Experimentally measured values of magnetic induction for a rectangular specimen. The lowest trace corresponds to a sensor very near the edge of the crystal and the others are successively nearer the centre with the top one about half way in. Data for increasing and decreasing fields have been averaged.

Figure 6 The entropy change at the phase transition in units of $k_B$ per flux line per layer as a function of temperature, calculated from the local field data reported in [1]. Graph (1) is calculated as in [1], graph (2) includes the factor of two relating the field change at the surface to the magnetisation, and graph (3) is as (2), but using $dH_m/dT_m$ instead of $dB_m/dT_m$ in the Clausius-Clapeyron equation.

Figure 7 The application of the Clausius-Clapeyron equation to the case of a phase transition which involves an intermediate state.